\documentclass[%
 reprint,
 amsmath,amssymb,
 aps,
]{revtex4-1}

\usepackage[dvipdfmx]{graphicx}
\usepackage{dcolumn}
\usepackage{bm}

\begin{document}

\preprint{}

\title{Strain analysis based on EAM and applications on \\ surface, vacancy, and boundary of Al}
\thanks{}%

\author{Shigeto R. Nishitani}
 \email{nishitani@kwansei.ac.jp}
 \homepage{https://ist.ksc.kwansei.ac.jp/~nishitani}
 \affiliation{%
Department of Informatics , Kwansei Gakuin University.}

\date{\today}

\begin{abstract}
  Stress or strain analysis for each atom around
  structural defects in a crystal is difficult.
  We propose a new analytical approach based on
  the eminent Embedding Atom Method(EAM) potential.
  We observe that the ratio $R$ between the repulsive
  and binding terms of the EAM is a definite measure
  for calculating the strain field of a single atom
  subject to an irregular coordination number.
  The determination of adequate potential parameters
  and their application to the calculation of the
  properties of surface, vacancy, and
  boundary of pure Al are shown.
\end{abstract}

\maketitle

\section{Introduction}
\label{sec:org2d4869d}
Geometrical phase analysis (GPA) is an innovation 
in the area of materials science, 
and has been improved over the past few decades\cite{gpa}.
This novel method allows us to describe the exact strain fields
that are present around dislocations 
by utilizing high resolution TEM images or 
atomistic simulations\cite{gpa_dislocation_nature, gpa_dislocation_zhao}.
When the coordination numbers of the atoms around a dislocation change,
it is not trivial to understand its compression or expansion behaviors.

The other classical method, the Voronoi polygon analysis\cite{voronoi},
is suitable in this case,
because it involves easier calculations for
changes in the coordination number.
In this method, a polygon is well defined by the surrounding atoms.  
However, for a vacancy or a surface, 
the volume of a specific atom can not be well defined.
In fact, the subsequent atoms might be too far away, 
or infinitely distant, due to these structural defects.

The atomistic simulations performed by calculations
based on first principles
provide reliable values for the positions of the atoms and the total energy of the system.
However, precise stress or energy value of each atom is hardly obtained.
Rigorous efforts have been dedicated to establishing the 
atomic size dividings of the total values based on the first principles calculations
\cite{shiihara_surface, shiihara_boundary}.
The difficulty of this approach is rooted in the ambiguity of
dividing the wave functions in the real space.

In order to avoid these difficulties, 
we propose a new analytical approach based on 
the eminent Embedding Atom Method(EAM) potential.
We propose that the ratio \(R\) between the repulsive and binding terms
of the EAM is a definite measure for calculating the strain field per atom
even with the altered coordination numbers.
In this paper, we first recall the EAM potential method,
especially with the applied first nearest neighbor model.
Secondly, we compare the structural energies and equilibrium distances of 
different perfect lattices 
and observe how the changes in coordination number affect the EAM potential.
Subsequently, we discuss the fitting of the EAM potential for
a suitable prediction of the strain field in different systems.
Finally, we observe a few examples of the strain field analysis of the EAM 
on the defect structures in pure aluminum, 
such as surface, vacancy, and boundary.

\section{Ratio between repulsive and bonding energies}
\label{sec:org2761417}
The embedded atom method (EAM) was developed for metals 
by Daw and Baskes\cite{DawBaskes}; 
this method essentially considers each metal atom to be embedded
within an electron density field generated by the surrounding atoms:
\begin{equation}
\label{eq:org9f0f628}
E_{\rm total} =E_{\rm repulsive} - E_{\rm bond}
= \sum \phi_{ij} - F(\rho) \, ,
\end{equation}
where \(\phi_{ij}\) is the simple pair repulsive interaction between \(i-j\) atoms, 
and \(\rho\) is the electron density of the field.  
When we take \(F(\rho) = \sqrt{\rho}\) 
and we express the electron density 
with the following 
equation defined by hopping integrals \(h(r_{ij})\),  
$$
\rho = \sum h(r_{ij})^2 \, ,
$$
we can obtain the Finnis-Sinclaire potential\cite{FinnisSinclaire};
this potential is equivalent to 
the one obtained by the second-moment models of Ducastelle \cite{Ducastelle}.
This second-moment approximation limits the identification of the differences 
in structural energy between hcp and fcc structures\cite[p.223]{David}.
Higher moment models of bond order potentials, including the angular dependence on bonding,
show different tendencies for the values of the structural energies 
depending on the dimension of the lattices \cite{bop_structure};
herein, the three-dimensional structures are always more stable 
than the two- or one-dimensional structures.
Among the three-dimensional structures, 
the dependence of cohesive energy on the equilibrium distance
between two atoms
primarily shows the same trend as that in the second moment models.
Because the targets in this paper are the defects in three dimensional lattices,
we settled to a second moment approximation.

We introduce two more constraints to simplify the analysis.
First, we select a simple exponential function 
for the pair and the hopping interactions as follows,
\begin{align}
\label{eq:org44999a4}
\phi(r_{ij}) & = a_0\exp(-p r_{ij}) \\
h(r_{ij}) & = b_0\exp(-q r_{ij}) 
\end{align}
where \(r_{ij}\) is the inter-atomic distance between \(i\) and \(j\) atoms.
The coefficients \(a_0\) and \(b_0\) are usually determined by the equilibrium condition.
The structural energy is simply determined by the \(p/q\) ratio,
where fcc structure is more stable than diamond structure for \(p/q>2\), and
diamond is more stable than fcc for \(p/q<2\) \cite{bop_structure}.

Second, we consider only the first nearest neighbor interaction.
We observe that the ratio between \(E_{\rm repulsive}\) and \(E_{\rm bond}\)
is a constant at its equilibrium distance and is independent of the coordination number \(n\).
When we consider a perfect lattice with the identical inter-atomic distance, 
the total energy per atom can be simply given as
\begin{align}
\label{eq:orga90c226}
E_{\rm total} &= \sum a_0 \exp(-p r) - \sqrt{\sum b_0^2 \exp(-2q r)} \nonumber \\
&= n a_0 \exp(-p r) - b_0 \sqrt{n \exp(-2q r)} \, .
\end{align}
The first derivative of distance \(r\) of this equation is as follows;
\begin{align}
\label{eq:orgd0fe5d2}
\frac{dE_{\rm total}}{dr} &= - np a_0 \exp(-p r) + qb_0 \sqrt{n \exp(-2q r)}
\end{align}
The solution of the note of this equation
gives the equilibrium distance \(r_n\) for a coordination number of \(n\), 
and gives the following relation;
\begin{equation}
\label{eq:org9c54e49}
\frac{qb_0}{pa_0\sqrt{n}} = \left(\exp(q-p)\right)^{r_n} .
\end{equation}
Calculating the ratio \(R\) between \(E_{\rm repulsive}\) and \(E_{\rm bond}\),
we obtain the following:
\begin{align}
\label{eq:org1726ebc}
R &= \frac{E_{\rm repulsive} }{E_{\rm bond}} 
=\frac{n a_0\exp(-p r_n)}{b_0 \sqrt{n \exp(-2q r_n)}} \\
&= \sqrt{n} \frac{a_0}{b_0} \left(\frac{\exp(q)}{\exp(p)}\right)^{r_n}
\end{align}
Now, substituting \(r_n\) from eq.(\ref{eq:org9c54e49}) will result in the following relation,
\begin{equation}
\label{eq:orga1f1b44}
R = \sqrt{n}\frac{a_0}{b_0} \frac{qb_0}{pa_0\sqrt{n}} = \frac{q}{p}
\end{equation}
which is independent of the coordination number.

This relation holds correctly for the perfect lattices, 
such as those cases when \(n=\) 12, 8, 6, and 4, 
which correspond to fcc, bcc, simple cubic and diamond structures, respectively.
It is also valid for imperfect lattices, 
such as those cases where \(n=\) 11, 10, 9, 7, 5, 3, 2, and 1.
Thus, the ratio \(R\) will be a measure of the strain field around a single atom;
this is true for any environment consisting of different coordination numbers, 
as shown in the next section in detail.

\section{Fitting to the first principles calculation results}
\label{sec:orgcf377aa}
Empirical potentials are fitted to the physical properties of real materials,
and one easy way is to use the first principles calculation results for this fitting procedure.  
The target of this study is to characterize the strain field 
for structures with different coordination numbers.
Thus, the structural energies obtained from the first principle calculations
are the starting point for the fitting of the EAM.

\begin{figure}[tbp]
\centering
\includegraphics[width=8.5cm]{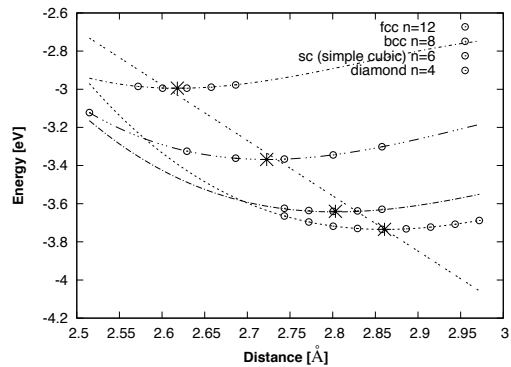}
\caption{\label{fig:org4424d98}
Structural energy and minimum positions for the perfect lattices of Al obtained by the first principles calculations.}
\end{figure}

Figure \ref{fig:org4424d98} shows the dependence of structural energy 
on the inter-atomic distance,
and the values corresponding to the minimum positions 
for the perfect lattices of Al obtained by the first principles calculations
with the coordination number \(n\) = 12, 8, 6 and 4 for fcc, bcc, sc (simple cubic) and diamond, respectively.
The calculation conditions are as follows:
To begin with, the first principles calculations are performed 
by VASP (Vienna ab initio simulation package)\cite{KresseHafner93},
implementing PAW (Projector Augmented Wave)\cite{KresseJoubert99} 
and GGA (Generalized Gradient Approximation)\cite{PerdewWang92} 
for the calculations of pseudo potentials, 
and the default energy cut off for the plane wave.
The \(k\) -point mesh used in each calculation was automatically generated by VASP,
because the cell sizes depend on the perfect lattices.
For this automatic generation, the length parameter defined in VASP
was set to 100 after checking for energy convergence. 

The first principles calculation results show the clear tendency of 
a decrease in inter-atomic distance with decreasing coordination numbers\cite{bop_structure}.
The minimum values for distance and energy show a nearly linear relation.
The straight line plotted in Fig. \ref{fig:org4424d98} is calculated
by the least square fit passing through the minimum position of the fcc lattice.
In the bcc lattice, the second nearest neighbors are the closest
in comparison to the other structures;
however, this feature is not obvious from this plot and will be discussed later.

\begin{figure}[tbp]
\centering
\includegraphics[width=8.5cm]{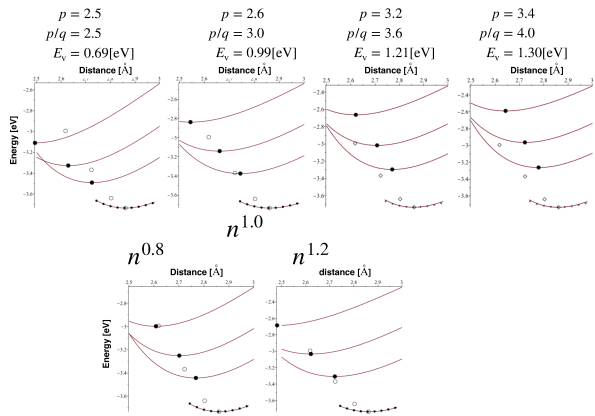}
\caption{\label{fig:org6fa1b04}
\(p/q\) dependency of the equilibrium points on the structural energy (open circles), with the first principles calculations (solid circles).}
\end{figure}

The explicit expression of the EAM potentials is shown in eq.(\ref{eq:orga90c226})
where \(n\) denotes the coordination number, 
and \(a_0\) and \(b_0\) are determined from the equilibrium condition,
the cohesive energy and the equilibrium distance.
\(p\) and \(q\) define the distance dependence of the potentials, and are
determined from the curvature of the structural energy function
and correspond to the bulk modulus.
The ratio \(p/q\) determines the stable points of the structures 
as shown in Figs. \ref{fig:org6fa1b04}.

Firstly, we set \(p/q\) to be 2.5
and \(p\) was determined by the bulk modulus of the fcc structure.
Based on these values, the stable points for bcc, sc and diamond can be determined.
The slope of the line connecting the stable points will increase
as the \(p/q\) value increases.
The vacancy formation energy is also related to the \(p/q\) values,
and this relation is illustrated in Figs. \ref{fig:org6fa1b04},
where an increase in the \(p/q\) ratio results in an increase in the vacancy formation energy.

Taking a power of the coordination number, such as \(n^{0.8}\) or \(n^{1.2}\), 
the stable points can be fitted more closely to the results obtained from the first principles.
From the technical point of view of simulations,
this effect is not calculated from the surrounding sites, 
but is usually included through the embedding function \(F(\rho)\) in Eq.(\ref{eq:org9f0f628}).
Because the change in this function alters the final result of the dependence of coordination on
the ratio between bonding and repulsive energies
that were obtained, as stated in the previous section,
we don't adopt a power of the coordination number.

\begin{figure}[tbp]
\centering
\includegraphics[width=8.5cm]{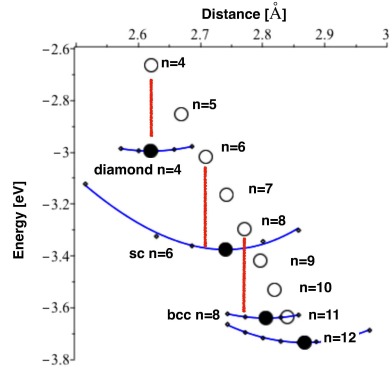}
\caption{\label{fig:org36333bc}
The equilibrium point change for different coordination numbers (open circles), with the structural energies obtained by first principles calculations (solid circles).}
\end{figure}

The parameters, \(p=3.2\) and \(p/q=3.6\), are the adopted values
used in the rest of the paper,
whose equilibrium points for different coordination numbers are shown in 
Fig.\ref{fig:org36333bc}.
The equilibrium points are not close to those of first principles results
shown in the same figure.
The reason for the selection of these parameter values is explained as follows.

Table \ref{tab:org48c2cb4} shows the comparison of the structural energy and,
inter-atomic distances obtained with the EAM prediction with the selected parameters,
and the first principles calculations.
For the case of the diamond lattice, 
the energy of -2.6635 eV is obtained by EAM;
this value is significantly different from -2.9943 eV,
which is obtained by the first principles calculations,
while the inter-atomic distance from the EAM, 2.6207 \AA{}, is close to the first principles value of 2.6179 \AA{}. 
The probability of this trend between discrepancies and similarities 
is the same for the other structures as well.
The minimum point for bcc structure is significantly different from those for the other structures;
this is reasonable because the EAM potential neglects the 
interactions farther than the first nearest neighbors,
which are relevant and should be included for bcc structures.
Because the surrounding atoms around mono-vacancy in fcc structure have the coordination number of 11,
the vacancy formation energies should be related to those systems with coordination number of 11, 
and are also much different.
Because the aim of applying the EAM potential in this research
is not for energy assessment but for strain analysis,
we therefore selected the parameters which better reproduce accurate inter-atomic distances
rather than energy values.

\begin{table}[tbp]
\caption{\label{tab:org48c2cb4}
Comparison of the energy and distance values between predictions by the EAM and values by the first principles calculations.}
\centering
\begin{tabular}{llrr}
\hline
structure & energy and & EAM potential & first principle\\
 & distance &  & \\
\hline
bcc & energy         [eV] & -3.2967 & -3.6397\\
 & distance           [\AA{}] & 2.7707 & 2.8032\\
Simple cubic & energy[eV] & -3.0174 & -3.3681\\
 & distance           [\AA{}] & 2.7084 & 2.7226\\
Diamond & energy     [eV] & -2.6635 & -2.9943\\
 & distance           [\AA{}] & 2.6207 & 2.6179\\
Vacancy & energy [eV] & 1.21 & 0.7\\
formation &  &  & \\
\hline
\end{tabular}
\end{table}

\begin{figure}[tbp]
\centering
\includegraphics[width=8.5cm]{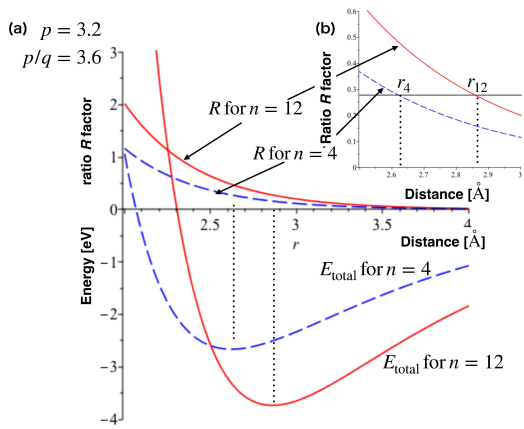}
\caption{\label{fig:org677c6f9}
\(E_{\rm total}\) and \(R\) dependency on distance for \(n=12\) (fcc) and \(4\) (diamond). The inset figure (b) gives a closer view around the equilibrium distances and \(R\) dependencies.}
\end{figure}

Figures \ref{fig:org677c6f9} shows the
\(E_{\rm total}\) and \(R\) dependency on the distance 
for coordination numbers \(n\) =12 (fcc) and 4 (diamond).
The inset figure (b) gives a
closer view around the equilibrium distances and \(R\) trends.
At each equilibrium distance of \(r_{12}\) = 2.8584 and \(r_{4}\) = 2.6207 \AA{}
for fcc and diamond, respectively,
the ratios \(R\) show the same value of 0.27778, which is the inverse of \(p/q\).
At the minimum distance for the other coordination numbers,
\(R\) shows the same value as expected by the theoretical derivations.  
Furthermore, the distance dependence of \(R\) for each coordination number
shows a monotonous dependency to the inter-atomic distance.
Thus \(R\) is a good measure for the strain fields;
a compressing field is observed for \(R>0.27778\) 
and an expanding field is observed for \(R<0.27778\).
We will see the applications of this measure around some lattice defects
in the next section.

\section{Applications on defects}
\label{sec:org003a203}
The defects we focus on in this section are surface, vacancy, and boundary, in pure aluminum.
The configurations for these defects are obtained by the automatic ionic relaxations of VASP 
with the parameters described in the previous section.

\subsection{Surface relaxations}
\label{sec:orga68b3ab}
\begin{figure}[tbp]
\centering
\includegraphics[width=8.5cm]{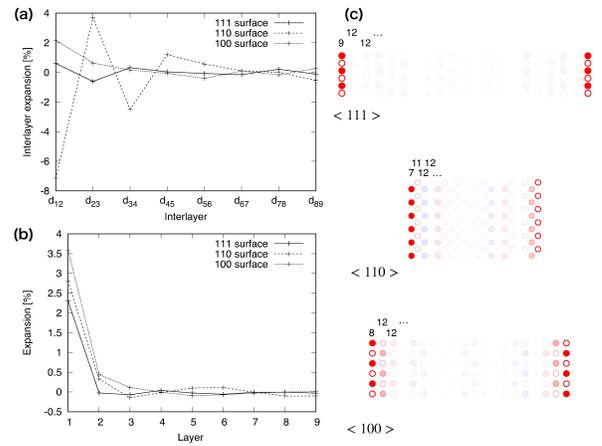}
\caption{\label{fig:org9fefca2}
Strain field diagrams for surfaces displaying \(\left(100\right)\), \(\left(110\right)\) and \(\left(111\right)\) surfaces; (a) inter-layer spacing, (b) strain field analysis and (c) strain field representation using colors of defferent intensities.}
\end{figure}

Figures \ref{fig:org9fefca2} shows the relaxation trends
for \(\left(100\right)\), \(\left(110\right)\) and \(\left(111\right)\) surfaces.  
Figure \ref{fig:org9fefca2}(a) shows a traditional plot of the inter-layer spacing
dependencies to the distance of the layers from the surface.  
We can compare them to the experimental and calculated results summarized by
Rodrigeuz \emph{et al}\cite{Rodriguez}.  The inter-layer spacing is depicted
as a \(\%\) of the inter-layer spacing of a perfect lattice depending on
the marks of the first-second, second-third inter-layer spacing
\(d_{12}\), \(d_{23}\) and so on.  A typical behavior of Al surface
relaxation is observed in the \(\left(110\right)\) surface, where a
10\(\%\) contraction of the first inter-layer and a \(5\%\) expansion of
the second inter-layer are observed both in experiments and
calculations.  Conversely, the first inter-layer of the
\(\left(100\right)\) and \(\left(111\right)\) surfaces in experiments and
calculations show the scattered results of small positive or negative
values.  The results of our study show the expansions in both
surfaces; however, the values are small.

Comparing to the variable trends in the inter-layer spacing plots,
our strain field analysis shows an identical behavior
on the relaxations of the different surfaces of Al.
Figure \ref{fig:org9fefca2} (b) shows 
the expanding distances calculated from the relations displayed
in Figs.\ref{fig:org677c6f9} using the value \(R\) from the EAM.
All atoms in the first layer expand about 2 to 4\(\%\) in all surfaces,
while the atoms in the second layer behave differently for different surfaces.
Although the atoms of the second layer in the \(\left(111\right)\) surface shows 
approximately no change in the absence of the strain field,
those in the \(\left(100\right)\) and \(\left(110\right)\) surfaces remain in the expanding positions.
The oscillating behavior shows a different tendency in \(\left(111\right)\),  
\(\left(100\right)\) and \(\left(110\right)\) surfaces.
In the \(\left(111\right)\) surface, the oscillation of the strain field drops quickly
after the first layer, and very small oscillation is observed in the third layer.
Alternatively, for the \(\left(100\right)\) and \(\left(110\right)\) surfaces, the oscillations is preserved until
the fourth and fifth layers.

These significantly different dependencies between the strain field and
the inter-layer distances, as seen in Figs.\ref{fig:org9fefca2}(a) and (b),
are explained by the differences in coordination number.
The coordination numbers of each layer are shown in the small characters
on the left side of Fig.\ref{fig:org9fefca2}(c); (9,12,12,\(\ldots\) ),  
(7,11,12,12, \(\ldots\) ), and (8,12,12,\(\ldots\) ) 
for \(\left(111\right)\), \(\left(110\right)\), and \(\left(100\right)\) surfaces, respectively.
We set the cut off distance between the first and the second nearest neighbor sites,
which is clearly determined for each surface.
When we change the measure from the distance to the ratio \(R\),
the expansion or compression magnitudes begin to depend on the coordination number.
The large compression of the first inter-layer of the \(\left(110\right)\) surface
becomes an expansion for the atoms with a coordination number of 7.
The oscillatory behavior observed in the inter-layer distances damps quickly;
the expansion and compression are only observed in those layers
with the coordination numbers different from 12 or just below. 

The three plots in Figs.\ref{fig:org9fefca2}(c) show
the color coded representations of the expansion-compression plots
highliting the magnitude for every atom;
red and blue represents expansion and compression, respectively,
and the brightness of the color represents the magnitude of these behaviors.
The open and closed circles represent the upper and lower layers 
along the direction perpendicular to the sheet.
The \(\left(111\right)\) surface shows a bright red color in the first layer, 
whereas it shows a very pale colors for the atoms below the second layer.
The \(\left(110\right)\) and \(\left(100\right)\) surfaces show red colored atoms 
in the second layer also,
whereas they show very pale colors below the third layer.
These representations are sufficient to grasp the expansion-compression field
and its magnitude.

The values of the strain can be compared with the stresses 
obtained by the first principles analysis
of Shiihara \emph{et al}\cite{shiihara_surface}.
Their estimation of the stresses on the \(\left(111\right)\) surface of Al 
comprises a large expansion for the first layer,
which is consistent with the results of this study.
On the other hand, for the second layer,
they predicted an in-plane compression,
and its magnitude is about one third of that of the first layer.
This oscillatory behavior in stress is rooted on the difficulty
of the dividing of the spaces around the different coordination numbers.

\subsection{Vacancy relaxations}
\label{sec:org148fffe}
\begin{figure}[tbp]
\centering
\includegraphics[width=8.5cm]{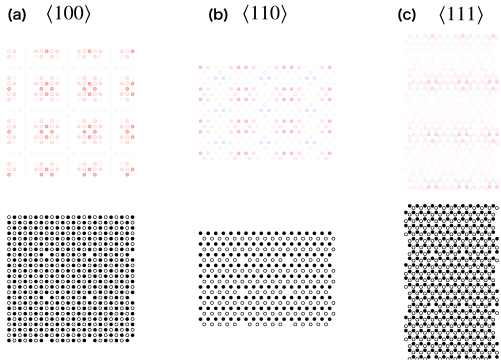}
\caption{\label{fig:org301262a}
Visual display of the strain field around vacancy.}
\end{figure}

Figures \ref{fig:org301262a} show visual representations of 
the strain field for the vacancy.  
Each column shows the cutting direction, 
and the first and the second layers intersecting vacancies are shown with
the open and closed circles, respectively.
The first nearest atoms around the vacancy show an expanding behavior
as same as in the surface.
An oscillatory behavior, on the other hand, 
cannot be observed in any sectioning planes around the vacancy.
This result is consistent with the results obtained from the first principles
calculation\cite{shiihara_boundary};
the first site around the vacancy shows a large expanding stress,
however, the other sites show nearly zero stress.

\subsection{Boundary relaxations}
\label{sec:org4f6d979}
\begin{figure}[tbp]
\centering
\includegraphics[width=8.5cm]{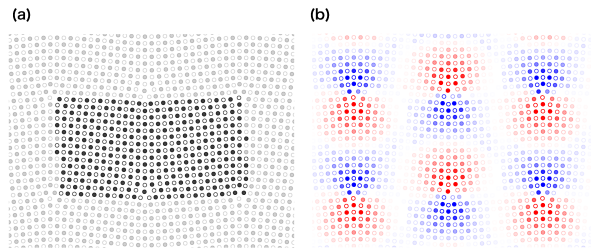}
\caption{\label{fig:org611ae57}
Visual representaion of the strain field display for boundaries, (a) for a relaxed atom configuration and (b) a color coded strain field representation.}
\end{figure}

Figures \ref{fig:org611ae57} show the strain fields around 
the \(\left<100\right>\) symmetric tilt grain boundary with an angle of 7.63\(^\circ\).
Figure \ref{fig:org611ae57}(a) shows a unit cell and its surroundings with
high and low contrasts, respectively.
The boundaries are located vertically on the center and
on both sides of the unit cell for which the periodic conditions are applied.
The open and closed circles represent the upper and lower
layers along \(\left[001\right]\) axis.
We can recognize the dislocations 
at equidistant positions along the boundaries.

Figure \ref{fig:org611ae57}(b) is a visual representation of the strain 
using \(R\) in the EAM analysis, where the color and color brightness of each site
mean the same as that in Figs.\ref{fig:org9fefca2}(c) and Figs.\ref{fig:org301262a}.
The region between the boundaries shows a pale color,
which indicates small strain as for a perfect lattice.
On the other hand, boundary regions are represented with an intense color.
The centers of the blue and red pairs are located at the dislocations
and the round shapes of these cells show
the typical shape of strain fields around dislocations
of the Peierls–Nabarro dislocation model \cite{gpa_dislocation_zhao}.

\begin{figure}[tbp]
\centering
\includegraphics[width=8.5cm]{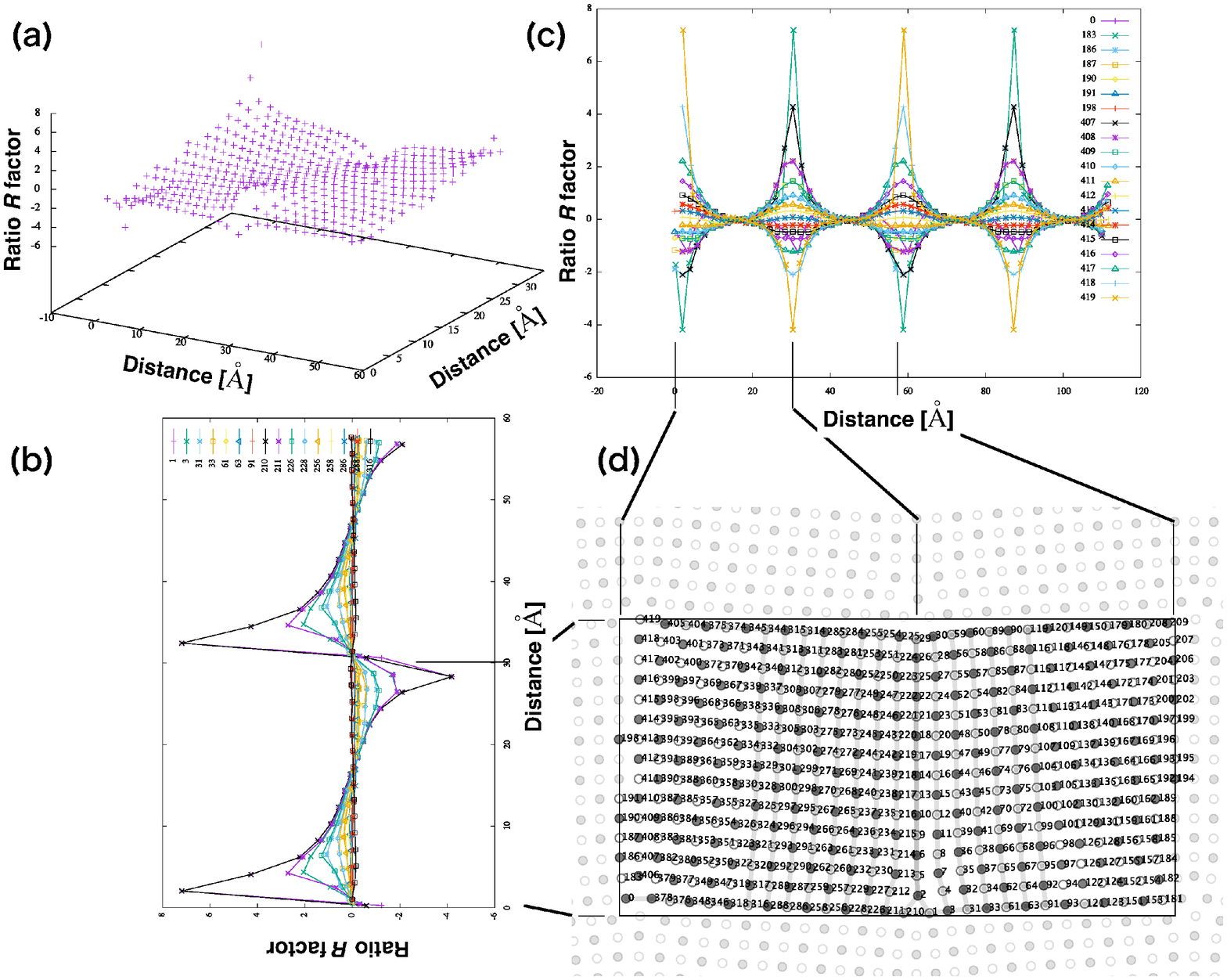}
\caption{\label{fig:org8f6dcd3}
(a) 3d, (b) vertical and (c) lateral plots of the strain field values calculated from the ratio \(R\) for boundaries.  (d) shows the indexing for the atoms.}
\end{figure}
Figures \ref{fig:org8f6dcd3} shows (a) 3d, (b) vertical and (c) lateral plots
of the strain field values.
The 3d parse view of Fig.\ref{fig:org8f6dcd3}(a) 
shows a unit cell,
where large deviations from the perfect lattice are observed around the dislocations
located at the boundary.
The vertical view along the boundary of Fig.\ref{fig:org8f6dcd3}(b) 
shows periodic peak and valley pairs.
Following the atom indexes as shown in Fig.\ref{fig:org8f6dcd3}(d),
No.1 and No.210 are located at the center bottom of the unit cell.
As shown in Fig.\ref{fig:org8f6dcd3}(b),
the lines pass through the maximum value of the expanding field at No.2,
and decrease to the minimum value, indicating a very large compression at No.29.
When we observe the lateral plots in Fig.\ref{fig:org8f6dcd3}(c),
No.419 is located at the right end of the upper most line,
and shows the largest expanding strain due to the surrounding configurations.
When the line moves right to the center of the unit cell,
the strain changes from expansion to compression,
and hits a minimum at No.29.
We can observe a smooth shifting of the strain field
with changes in the coordination number;
for No.2, the dislocation core has \(n=10\),
No.1 and No.210 also have \(n=10\), 
whereas No.3 and No.211 have \(n=11\).
No.4 and No.213 show different coordination numbers of 10 and 12, respectively,
due to the slight asymmetry of the relaxed configuration.
However, they show nearly identical values for the strain field.
This means that the slight difference of the coordination number
does not affect significantly on the strain field values calculated from 
the \(R\) factor of the EAM.

\section{conclusions}
\label{sec:orgd1c2805}
We have derived a measure of the strain field from EAM potentials.
The ratio \(R\) between the repulsive energy and
the bonding energy for single atoms shows the same value at
the minimum point for different coordination numbers \(n\);
these values were calculated under some approximations applied to the EAM potential
including using the square root function of the electron density,
the exponential dependency on the repulsive and hopping interactions,
and for the calculation of the first nearest neighbor interactions.
We have shown the applications of this novel method on
the surface, vacancy, and boundary of pure Al.
For each atom around these defects, having altered coordination numbers,
the expanding or compressing strain calculated from the \(R\) factor of the EAM
shows the smooth change, which is expected from continuum mechanics.

\section{Acknowledgement}
\label{sec:org9bf1229}
The author thanks to Mr. Shunji Nishitani, who showed the derivation
of Eq.(\ref{eq:orga1f1b44}) for the first time.
This work was carried out using 
the JFRS-1 supercomputer system at Computational Simulation Centre of 
International Fusion Energy Research Centre (IFERC-CSC) 
in Rokkasho Fusion Institute of QST (Aomori, Japan).

\bibliographystyle{ieeetr}
\bibliography{eam_analysis}

\begin{thebibliography}{10}

\bibitem{gpa}
M.~H\"{y}tch, E.~Snoeck, and R.~Kilaas, ``Quantitative measurement of
  displacement and strain fields from hrtem micrographs,'' {\em
  Ultramicroscopy}, vol.~74, pp.~131--146, 1998.

\bibitem{gpa_dislocation_nature}
M.~J. H\"{y}tch, J.-L. Putaux, and J.-M. P\'{e}nisson, ``Measurement of the
  displacement field of dislocations to 0.03 \aa by electron microscopy,'' {\em
  Nature}, vol.~423, pp.~270--273, 2003.

\bibitem{gpa_dislocation_zhao}
Z.~Dong and C.~Zhao, ``Measurement of strain fields in an edge dislocation,''
  {\em Physica B}, vol.~405, pp.~171--174, 2010.

\bibitem{voronoi}
G.~Voronoi, ``Nouvelles applications des param^^c3^^a8tres continus ^^c3^^a0 la
  th^^c3^^a9orie des formes quadratiques. premier m^^c3^^a9moire. sur quelques
  propri^^c3^^a9t^^c3^^a9s des formes quadratiques positives parfaites,'' {\em
  J. reine angew. Math.}, vol.~133, pp.~97--178, 1908.

\bibitem{shiihara_surface}
Y.~Shiihara, M.~Kohyama, and S.~Ishibashi, ``Ab initio local stress and its
  application to al (111) surfaces,'' {\em Phys. Rev. B}, vol.~81, p.~075441,
  2010.

\bibitem{shiihara_boundary}
H.~Wang, M.~Kohyama, S.~Tanaka1, and Y.~Shiihara, ``Ab initio local energy and
  local stress: application to tilt and twist grain boundaries in cu and al,''
  {\em J. Phys.: Condens. Matter}, vol.~25, p.~305006, 2013.

\bibitem{DawBaskes}
M.~S. Daw and M.~I. Baskes, ``Embedded-atom method: Derivation and application
  to impurities, surfaces, and other defects in metals,'' {\em Phys. Rev. B},
  vol.~29, p.~6443, 1984.

\bibitem{FinnisSinclaire}
M.~W. Finnis and J.~E. Sinclair, ``A simple empirical n-body potential for
  transition metals,'' {\em Phil. Mag. A}, vol.~50, pp.~45--55, 1984.

\bibitem{Ducastelle}
F.~Ducastelle, ``Modules ^^c3^^a9lastiques des m^^c3^^a9taux de transition,''
  {\em J. de Physique}, vol.~31, p.~11, 1970.

\bibitem{David}
D.~G. Pettifor, {\em Bonding and structure of molecules and solids}.
\newblock Clarendon Press(Oxford), 1995.

\bibitem{bop_structure}
S.~R. Nishitani, P.~Alinaghian, C.~Hausleitner, and D.~G. Pettifor, ``Angularly
  dependent embedding potentials and structural prediction,'' {\em Phil. Mag.
  Lett.}, vol.~69, pp.~177--184, 1994.

\bibitem{KresseHafner93}
G.~Kresse and J.~Hafner, ``Ab initio molecular dynamics for liquid metals,''
  {\em Phys. Rev. B}, vol.~47, pp.~558--61, 1993.

\bibitem{KresseJoubert99}
G.~Kresse and D.~Joubert, ``From ultrasoft pseudopotentials to the projector
  augmented-wave method,'' {\em Phys. Rev. B}, vol.~59, pp.~1758--75, 1999.

\bibitem{PerdewWang92}
J.~P. Perdew and Y.~Wang, ``Accurate and simple analytic representation of the
  electron-gas correlation energy,'' {\em Phys. Rev. B}, vol.~45, pp.~13244--9,
  1992.

\bibitem{Rodriguez}
A.~M. Rodr\'{i}guez, G.~Bozzolo, and J.~Ferrante, ``Multilayer relaxation and
  surface energies of fcc and bcc metals using equivalent crystal theory,''
  {\em Surf. Sci.}, vol.~289, pp.~100--126, 1993.

\end{thebibliography}
\end{document}